\documentclass{cernrep}
\usepackage{texnames}
\usepackage[T1]{fontenc}
\usepackage{varwidth}
\usepackage{xcolor}

\begin{document}

\title{Beam Loss Monitors at LHC}
\author{B. Dehning}
\institute{CERN, Geneva, Switzerland}
\maketitle

\begin{abstract}
One of the main functions of the LHC beam loss measurement system is the protection of equipment against  damage caused by impacting particles creating secondary showers and their energy dissipation in the matter. Reliability requirements are scaled according to the acceptable consequences and the frequency of particle impact events on equipment. Increasing reliability often leads to more complex systems. The downside of complexity is a reduction of availability; therefore, an optimum has to be found for these conflicting require\-ments. A detailed review of selected concepts and solutions for the LHC system will be given to show approaches used in various parts of the system from the sensors, signal processing, and software implementations to the requirements for operation and documentation.\\\\
{\bfseries Keywords}\\
Machine protection; equipment protection; beam loss; dependability.
\end{abstract}
\section{Introduction}
After a LHC beam loss project study phase, a functional specification is compiled. The specification introduces the subject, first viewing the project globally by treating:
\begin{itemize}
        \item location of monitors;
        \item time response;
        \item dynamic range;
        \item safety and reliability requirements.
\end{itemize}

The safety and reliability requirements need to be discussed at the system level, to define the overall quantitative requirements. The time response, dynamic range, safety, and reliability requirements limit the choice of sensors and define the acquisition chain. With the knowledge obtained in the project study phase, the following choices are made:
\begin{itemize}
        \item sensor: ionization chamber;        \item acquisition chain: distributed system with local and independent beam inhibit functionality.
\end{itemize}

A more detailed treatment of the global safety and reliability requirements has been covered in study groups and thesis projects. The subjects treated include:
\begin{itemize}
        \item acquisition chain with:
        \begin{itemize}
                \item parallel and voting for safety and reliability requirements;
                \item radiation-tolerant electronics;
        \end{itemize}
        \item fail-safe system;
        \item data flow path;
        \item management of settings;
        \item functional tests;
        \item preventive actions;
        \item firmware updates;
        \item reliability software;
        \item human errors;
        \item documentation.
\end{itemize}

Several of these aspects will be discussed in this paper, and examples will be presented from the LHC beam loss monitoring system.
\section{Global beam loss measurement requirements}
\begin{figure}
   \centering
   \includegraphics[width=100mm]{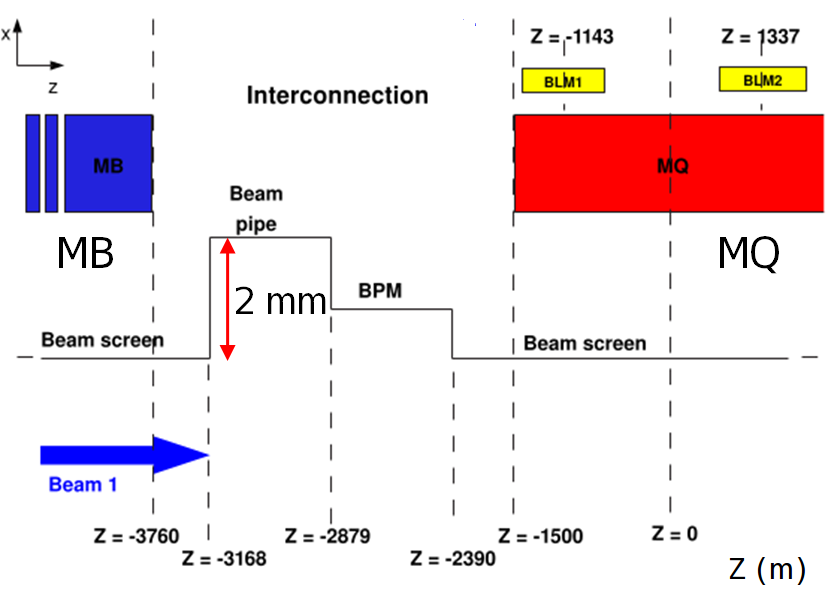}
   \caption{Loss location considerations: aperture between a LHC bending magnet (MB) and a quadrupole magnet (MQ). The change in aperture is mainly controlled by the connection bellow and the beam position monitor (BPM) location. BLM, beam loss monitor.}
   \label{BLM_at_LHCf011}
\end{figure}

\begin{figure}
   \centering
   \includegraphics[width=100mm]{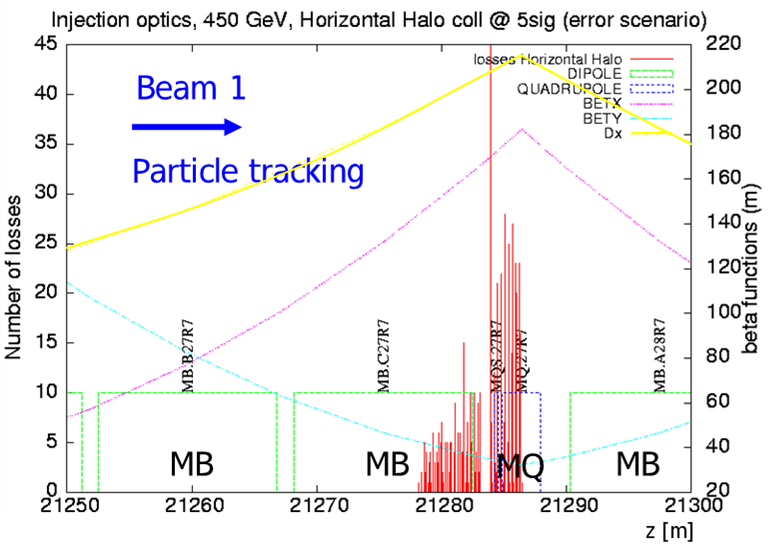}
   \caption{Number of lost protons and beta function values with schematic of LHC regular cell as function of the location along the lattice. MB, bending magnet; MQ, quenching magnet.}
   \label{BLM_at_LHCf012}
\end{figure}

For a beam loss protection system, the possible loss locations and therefore also the potential damage location are unknown parameters, to be addressed by particle tracking and particle shower simulations. In a second step, the optimal sensor locations are also determined by particle shower simulations. For the LHC, the considerations are illustrated in Fig.~\ref{BLM_at_LHCf011}. The electrodes of the beam position monitors are retracted to be shielded by the nearby vacuum chamber walls against particle impacts, which could create electrical charges on the electrodes and disturb the beam position measurement.
An aperture limitation results in a concentration of losses if off-orbit protons approach the aperture. At the LHC, this is the case for every transition between a bending and a quadrupole magnet. This can be visualized by the tracking simulation (Fig.~\ref{BLM_at_LHCf012}), resulting in a maximum at the beginning of the quadrupole magnet.
These loss locations are most probable, because:
\begin{itemize}
        \item the beta function, and therefore the beam size, is maximal;
        \item orbit bumps have a maximum at this location, because of the location of a dipole corrector magnet near to the quadrupole
magnet;
\item alignment errors are possible, causing an additional aperture limitation.
\end{itemize}

\begin{figure}
   \centering
   \includegraphics[width=100mm]{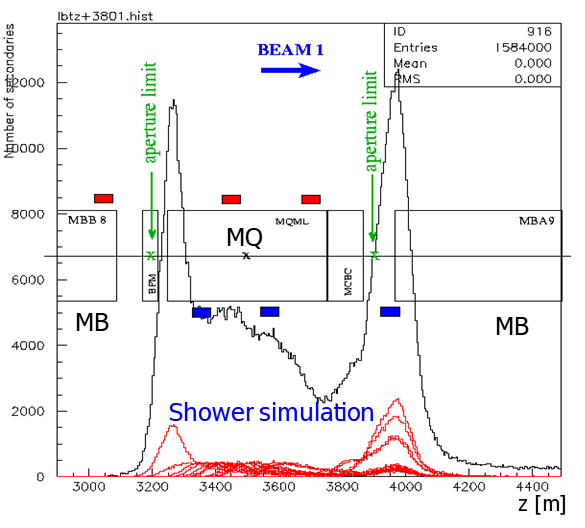}
   \caption{Number of secondary particles as function of location along the lattice. MB, bending magnet; MQ, quenching magnet.}
   \label{BLM_at_LHCf013}
\end{figure}

The shower particles initiated by lost protons can be best observed outside of the magnet yoke about a metre downstream of the proton impact location (see Fig.~\ref{BLM_at_LHCf013}).
A second maximum occurs at the downstream transition between the quadrupole and bending magnet, owing to the reduced material in the transition region.
To make use of the high particle signal, resulting in the lowest statistical measurement error, the ionization chambers are located at or near to particle shower maxima (see Fig.~\ref{BLM_at_LHCf013}, red and blue rectangular areas). A separation between the losses from beams 1 and 2 is given by the different locations of the shower particle maxima, owing to their opposite directions.

\begin{figure}
   \centering
   \includegraphics[width=120mm]{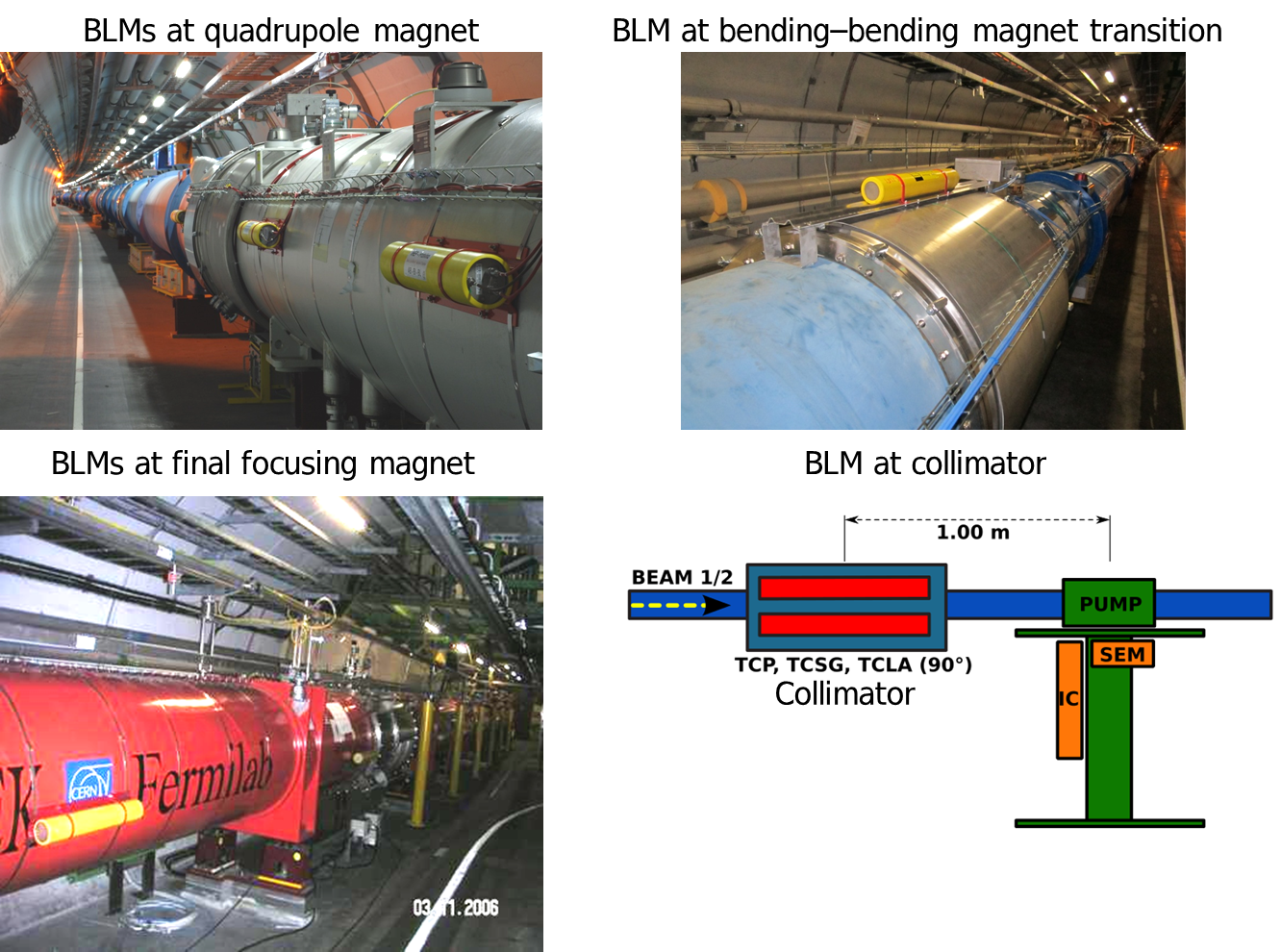}
   \caption{LHC tunnel photos with ionization chambers (yellow tubes) mounted on the outside of magnets and schematic of an ionization chamber near a collimator. BLM, beam loss monitor; IC, ionization chamber; SEM, secondary emission monitor.}
   \label{BLM_at_LHCf02}
\end{figure}

The LHC ionization chambers are cylindrical, with a sensitive volume of $\Unit{1.5}{l}$,  covered by a yellow insulating tube and are mounted on the outside of the magnets or near collimators (see Fig.~\ref{BLM_at_LHCf02}, bottom right, red and blue rectangular areas).

The limits of the time response and dynamic range requirements for LHC protection are mostly defined by the quench curves of the bending magnets. The quench levels of the magnets are orders of magnitude lower than the damage levels of the magnets. Magnet quenching is avoided, because of the gain in operational efficiency, by extracting the beam from the ring and therefore ending the deposition of heat in the coil before quenching can occur. In the case of a quench, the magnet coil is warmed up and the new cool down takes between 6 and $10\Uh$.
\begin{figure}
   \centering
   \includegraphics[width=120mm]{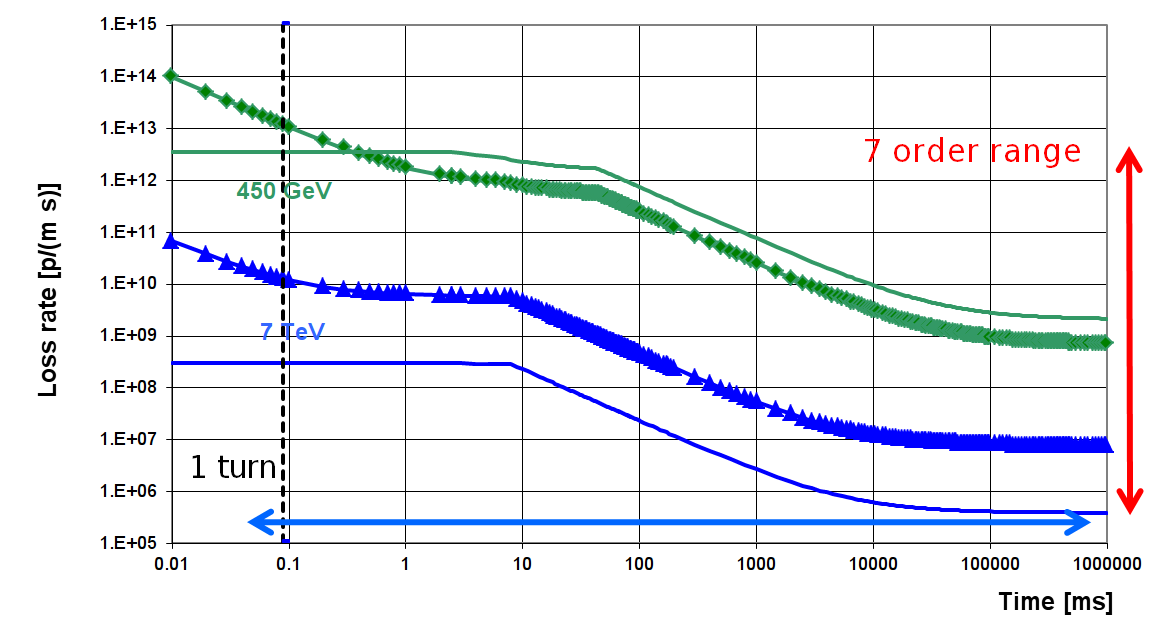}
   \caption{Proton density rate as function of loss duration. Different curves indicate the functional dependence for different energies and the defined observation range. Red arrow, required proton density rate dynamic; blue arrow, duration dynamic.}
     \label{BLM_at_LHCf03}
\end{figure}
The allowed particle loss rate (see Fig.~\ref{BLM_at_LHCf03}) in protons per metre per second is shown as the function of the loss duration. The characteristic superconducting magnet quench level curves are due to the quench margin of the superconducting cable filaments and the super fluid \EHe\ cooling of the cables and the whole magnet coil. For short duration losses, the quench level is about four orders of magnitude higher than for steady-state losses and for both LHC nominal beam energies, of $450\UGeV$ and $7\UTeV$,
an order of two variation is seen.

The time resolution of the loss measurement system of $40\Uus$ is given by the duration of the ex\-traction of the beam from the LHC, $89\Uus$, and some signal propagations and synchronization con\-sider\-ations. The maximum duration is given by the reach of the steady-state quench level at about $80\Us$ (see Fig.~\ref{BLM_at_LHCf03}, blue arrow).

The maximal signal value is defined by the crossing of the $89\Uus$ line and the quench level at $450\UGeV$. Owing to an optimization process for the LHC acquisition electronics, the value has been chosen a little lower (see Fig.~\ref{BLM_at_LHCf03}, vertical dashed black line ($89\Uus$) and thin green line). The lower limit of the dynamic range is given by the steady-state quench level for $7\UTeV$ and the need to observe losses, for accelerator tuning purposes, below the quench level (see Fig.~\ref{BLM_at_LHCf03}, thin blue line, $80\Us$). These considerations led to a required signal dynamic of over seven orders of magnitude (see Fig.~\ref{BLM_at_LHCf03}, red arrow). Operational experience required that the dynamic upper value be extended by two orders of magnitude for short-term losses in injection areas.

\section{Safety system design approach}
\begin{figure}
   \centering
   \includegraphics[width=120mm]{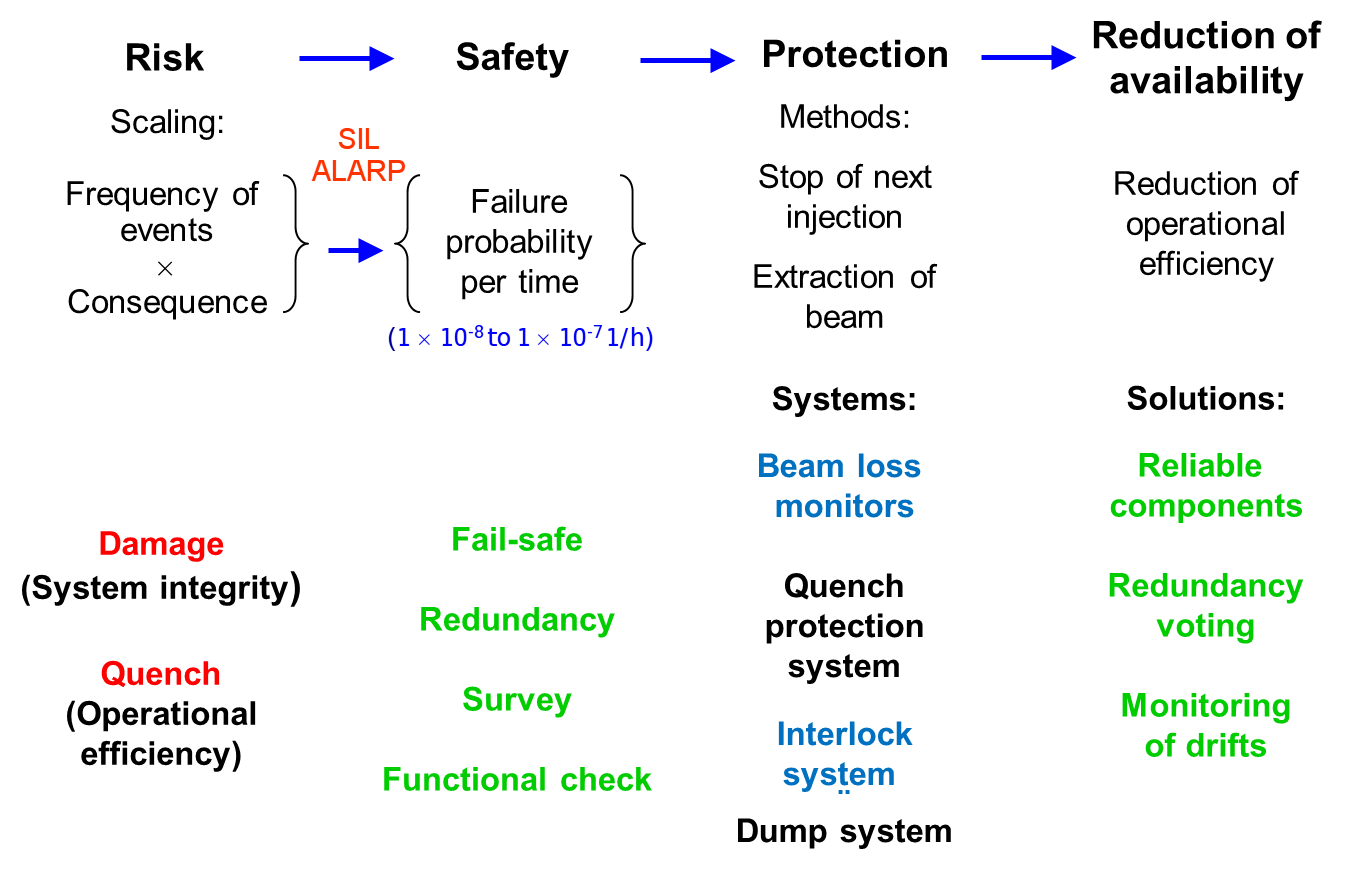}
   \caption{LHC protection system design approach (items in green are discussed in this paper). ALARP, as low as reasonably practicable; SIL, safety integrity level.}
    \label{BLM_at_LHCf1}
\end{figure}

All considerations start with the recognition that the probable frequency and probable magnitude of a non-conformal behaviour could lead to a damage of the system integrity. The combined likelihood of frequency and magnitude determines the risk for a certain system (see Fig.~\ref{BLM_at_LHCf1}, first column). The risk could be reduced by using a safety system providing protection, but increased complexity reduces the availability of the protected system (see Fig.~\ref{BLM_at_LHCf1}, first row).
To arrive at a quantitative demand for a safety level, the probable frequency of events and the probable magnitude of its consequence are utilized by the SIL (safety integrity level) approach~\cite{iec_2010} or the `as low as reasonably practicable' (ALARP) approach.

For both approaches, a failure probability per time is estimated by calculating the risk of damage and the resulting downtime of the equipment~\cite{guaglio_reliability_2004}. A failure in the safety system itself should fall in a fail-safe state, with the consequence of reducing the operation efficiency. The main design criteria for the safety system are listed in the safety column of Fig.~\ref{BLM_at_LHCf1}: fail-safe, redundancy, survey, and functional check. The protection column  of Fig.~\ref{BLM_at_LHCf1} lists the methods for the protection of an accelerator: stop of next injection applicable for a one-path particle guiding system (linac, transfer line) and extraction of the beam for a multipath system (storage ring). The accelerator safety system consists of a beam loss measurement system, an interlock system, and a beam dump system. If superconducting magnets are used, some beam loss protection could also be provided by the quench protection system. The availability column  of Fig.~\ref{BLM_at_LHCf1} lists the means used in the design of the safety system to decrease the number of transitions of the system into the fail-safe state. The effect of the number of components added to a system to increase the probability of a safe operation results in a reduction in the availability of the system. This negative consequence of the safety-increasing elements is partially compensated  by the choice of reliable components, by redundancy, voting, and the monitoring of drifts of the safety system parameters.

\section{Failure probability and failure rate reduction}\label{sec:failureprobability}
\begin{figure}
   \centering
   \includegraphics[width=80mm]{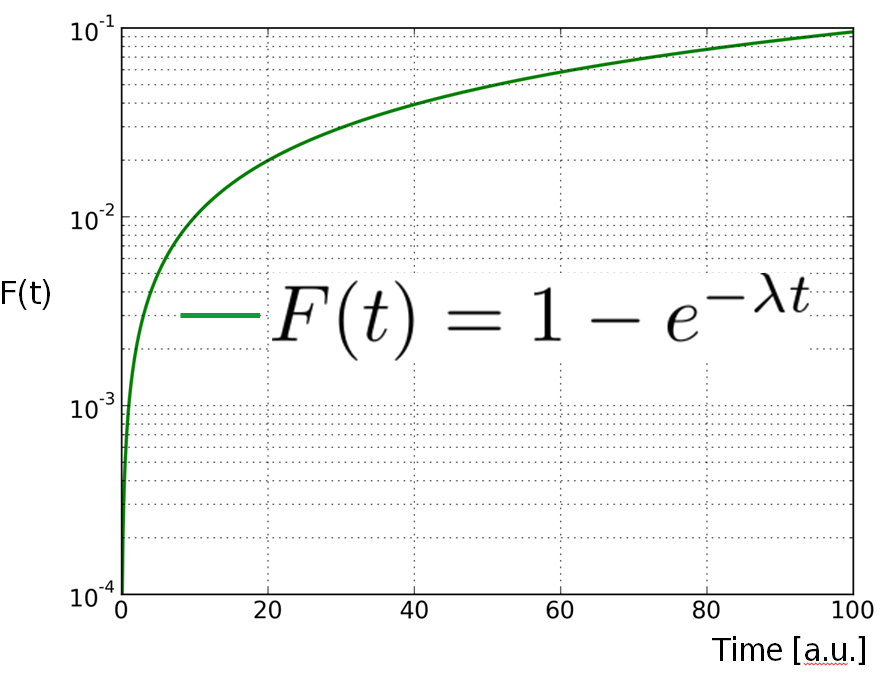}
   \caption{Exponential failure probability}
   \label{BLM_at_LHCf2}
\end{figure}

\begin{figure}
   \centering
   \includegraphics[width=80mm]{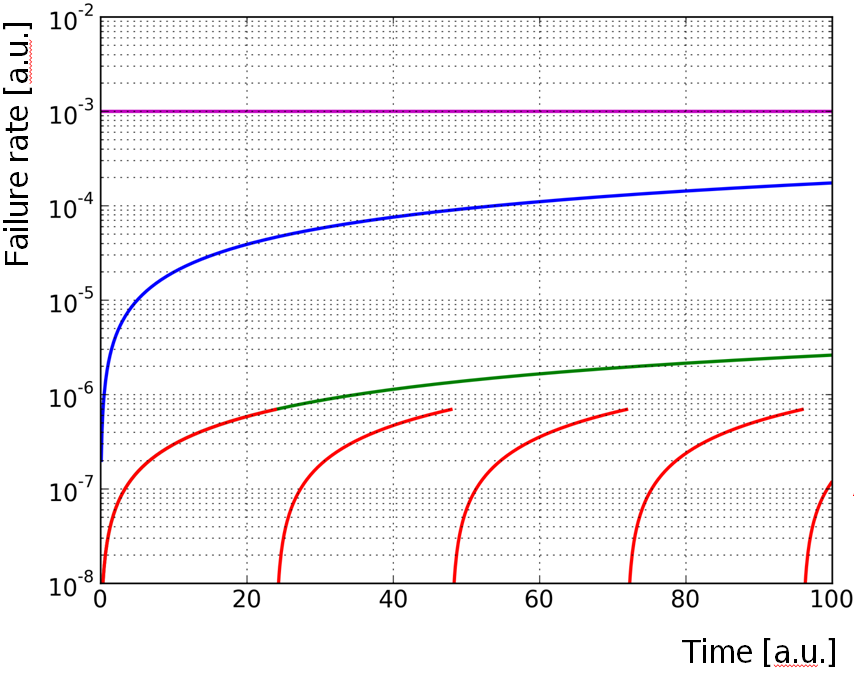}
   \caption{Failure rates of different systems as a function of time (arbitrary units). Magenta: single system. Blue: Two systems parallel. Green: Parallel systems with survey. Red: Parallel systems with survey  and with regular test.}
   \label{BLM_at_LHCf3}
\end{figure}

To illustrate the available means of increasing safety, the system's basic functional dependencies are discussed. An often-valid assumption is given by the exponential time dependence of the failure probability $F(t)$ (Fig.~\ref{BLM_at_LHCf2}). With increasing time, the probability of the occurrence of a failure in a system approaches~1. The failure rate, $\lambda$, is assumed to be time-independent (Fig.~\ref{BLM_at_LHCf3}, magenta curve). In a next step, two systems with the same functionality are assumed to be working in parallel, to allow redundant operation. The failure rate, $\lambda$, decreases drastically for short times, but finally approaches the failure rate of a single system (Fig.~\ref{BLM_at_LHCf3}, blue line).

It should be noted that the failure rate curve changes from time-independent to time-dependent behaviour. A further reduction in the failure rate could be reached by a survey of the system. With a system survey, some failure modes can be detected in advance
and a repair can be planned (see Fig.~\ref{BLM_at_LHCf3}, red and green line). This procedure results in a shift of the failure rate curve to lower values, which no longer approach the  infinite times of the single system rate. Another strong reduction could be reached if the system could be regarded as new after a certain time period. The failure rate curve shows the time dependence of the surveyed system in the period $ t_{0} = 0 $ to $ t = t_{1} $ repeated after every time period (see Fig.~\ref{BLM_at_LHCf3}, red lines). The conclusion that a system could be regarded as new after a certain time is justified if the system is subjected to a test. Functional tests will verify, on request, that the system has the defined functionality. In case of an internal system failure system, the very basic requirement is a fail-safe behaviour. Internal failure will not contribute to the unsafeness of the system but will contribute to its non-availability.

\section{Protection system overview}
As an example of a protection system, the CERN LHC beam loss monitoring (BLM) system will be used. The discussion will focus on protection, reliability, and availability aspects.

The main purpose of the BLM system is to convert particle shower information into electrical signals, which are then compared with limits. If the limits are exceeded, extraction of the LHC beam from the ring is initiated to stop the irradiation of equipment. In the case of the LHC, the protection function is often linked to the quench prevention of the superconducting magnets, since the threshold levels for beam extraction are lower (orders of magnitude) than for the damage protection of equipment~\cite{dehning_overview_2011}.

\begin{figure}
   \centering
   \includegraphics[width=160mm]{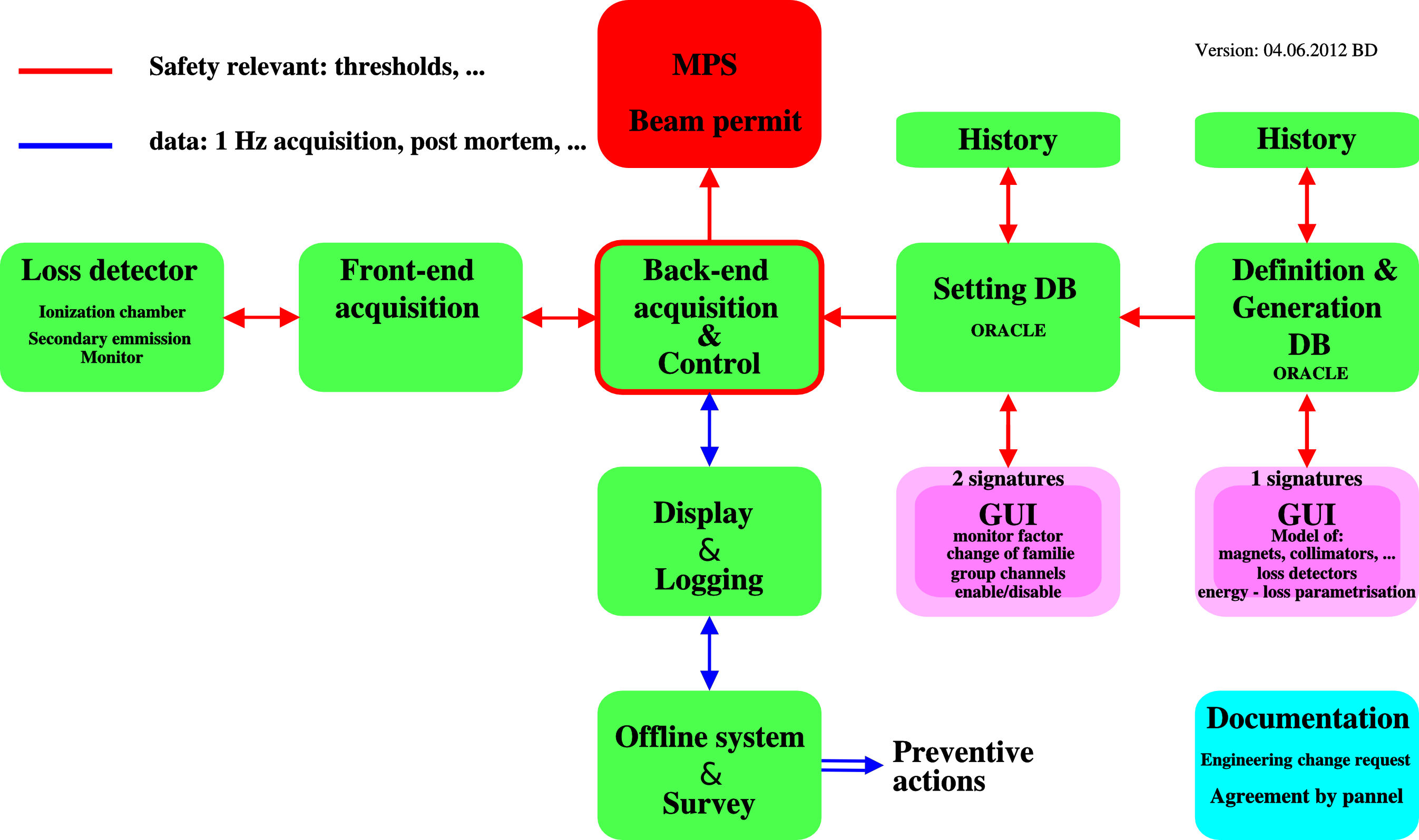}
      \caption{Information flow from the sensor up to the beam permit signal transmission. The red framed (back-end acquisition and control) unit is the local decision-making centre.}
    \label{BLM_at_LHCf4}
\end{figure}

The very first element of the protection system is the sensor that detects the irradiation of equipment.
The conversion of the particle shower magnitude is done by ionization chambers~\cite{stockner_beam_2006} or secondary emission detectors~\cite{kramer_design_2008} (see Fig.~\ref{BLM_at_LHCf4}, left block). The front-end acquisition electronics convert the analogue detector signal into a digital signal and transmit the signal to the back-end acquisition and control unit, which is the decision-making centre of the whole system. The measured signals arrive here and are compared with the limits. In addition, beam permit signals are generated (see Fig.~\ref{BLM_at_LHCf4}, red block), taking the information of the system settings (see Fig.~\ref{BLM_at_LHCf4}, right-hand blocks) into account. The measurement data and all setting information  are also distributed to the display and the logging databases (see Fig.~\ref{BLM_at_LHCf4}, bottom blocks) from this unit. The control functionality is linked to the survey and test functionality, which are discussed later.

\begin{figure}
   \centering
   \includegraphics[width=160mm]{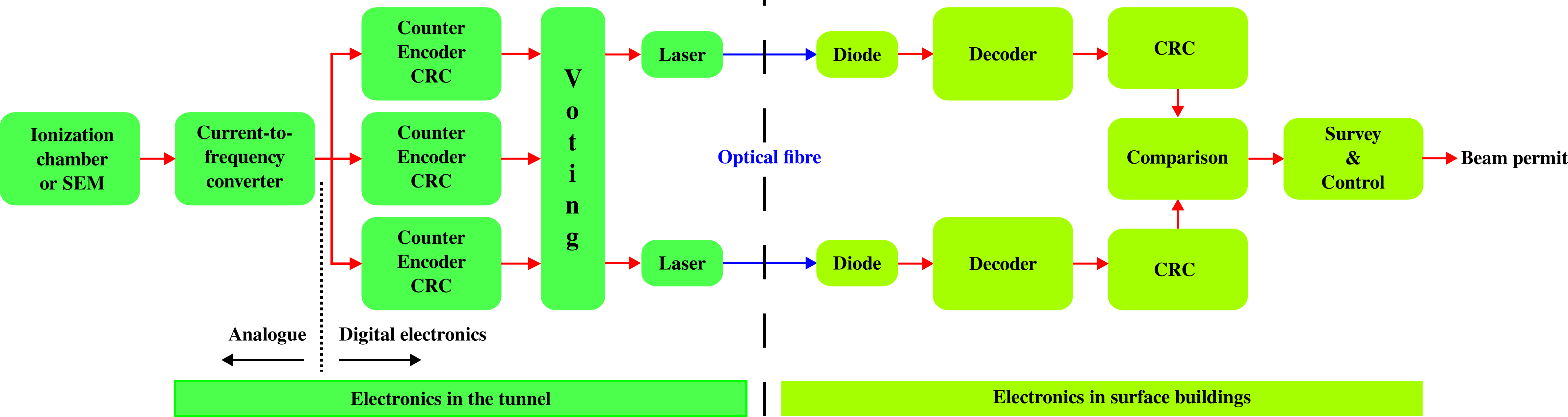}
   \caption{CERN LHC beam loss measurement and protection system: CRC, cyclic redundancy check}
      \label{BLM_at_LHCf5}
\end{figure}

In the LHC, ionization chambers~\cite{stockner_beam_2006} and secondary emission detectors~\cite{kramer_design_2008} are used. Their signals are digitized using a current-to-frequency converter~\cite{effinger_lhc_2006,effinger_single_2007} (see Fig.~\ref{BLM_at_LHCf5}, front-end acquisition unit in tunnel). Up to the end of the analogue signal chain, the signal is not redundant, because no technical solution has been found to the problem of splitting the detector signal while simultaneously allowing a large dynamic signal (nine orders of magnitude). To cope with this requirement for the analogue front-end unit, a low failure rate circuit concept has been chosen. To avoid the consequences of single event effects, and to increase the availability of a channel, the signal is trebled in the front-end logic. Two voting blocks are used to generate the signal transmitted over a redundant optical link. A redundant optical link has been chosen to increase the availability of the link, which is limited by the mean time between failures of the transmission laser.

The signals are decoded and cyclic redundancy checks (CRCs) are calculated for both signal chains (see Fig.~\ref{BLM_at_LHCf5}, back-end acquisition unit at the surface). At the front-end unit, CRCs are also calculated and transmitted, to enable the CRCs of each line and also the CRCs for both lines to be compared. This procedure ensures high reliability and also maximizes the availability of the data link~\cite{zamantzas_fpga_2006, zamantzas_real-time_2006}.

The effect of the implementation of redundancy and trebling in the data transmission and treatment and the verification of loss-free data transmission are listed in Table~\ref{table_reliability}. The most important technique for increasing the reliability of a system is given by a fail-safe design. In the case of an internal failure of a system, it should make the transition to a state that ensures the protection of the system. This could be done by assigning the active state to: `system is allowed to operate'. In case of an internal failure, \eg if no power is supplied, the state will switch to a passive state and the system will be protected.

\begin{table} \scriptsize
\centering
\caption{Procedure and techniques  to increase the reliability and availability of acquisition systems}
   \label{table_reliability}
   \vspace{1mm}
\begin{tabular}{l l c c}
\hline\hline
           & \textbf{Comment     position of monitor      }         & \textbf{Safety gain} & \textbf{Availability gain} \\  \hline
Fail-safe   & Active state = beam permit & Yes         & No \\
Voting     &                            & Yes         & Yes \\
Redundancy &                            & Yes         & Yes \\
CRC        & Cyclic redundancy check    & Yes         & No \\
\hline\hline
\end{tabular}
\end{table}

\section{Fault tree analysis}
The fault tree treatment of the system has been chosen to calculate, from the component level up to the system level, the damage risk, the false alarm, and the warning probability~\cite{guaglio_reliability_2005}, taking into account the component failure, repair and inspection rates.

\begin{figure}
   \centering
   \includegraphics[width=100mm]{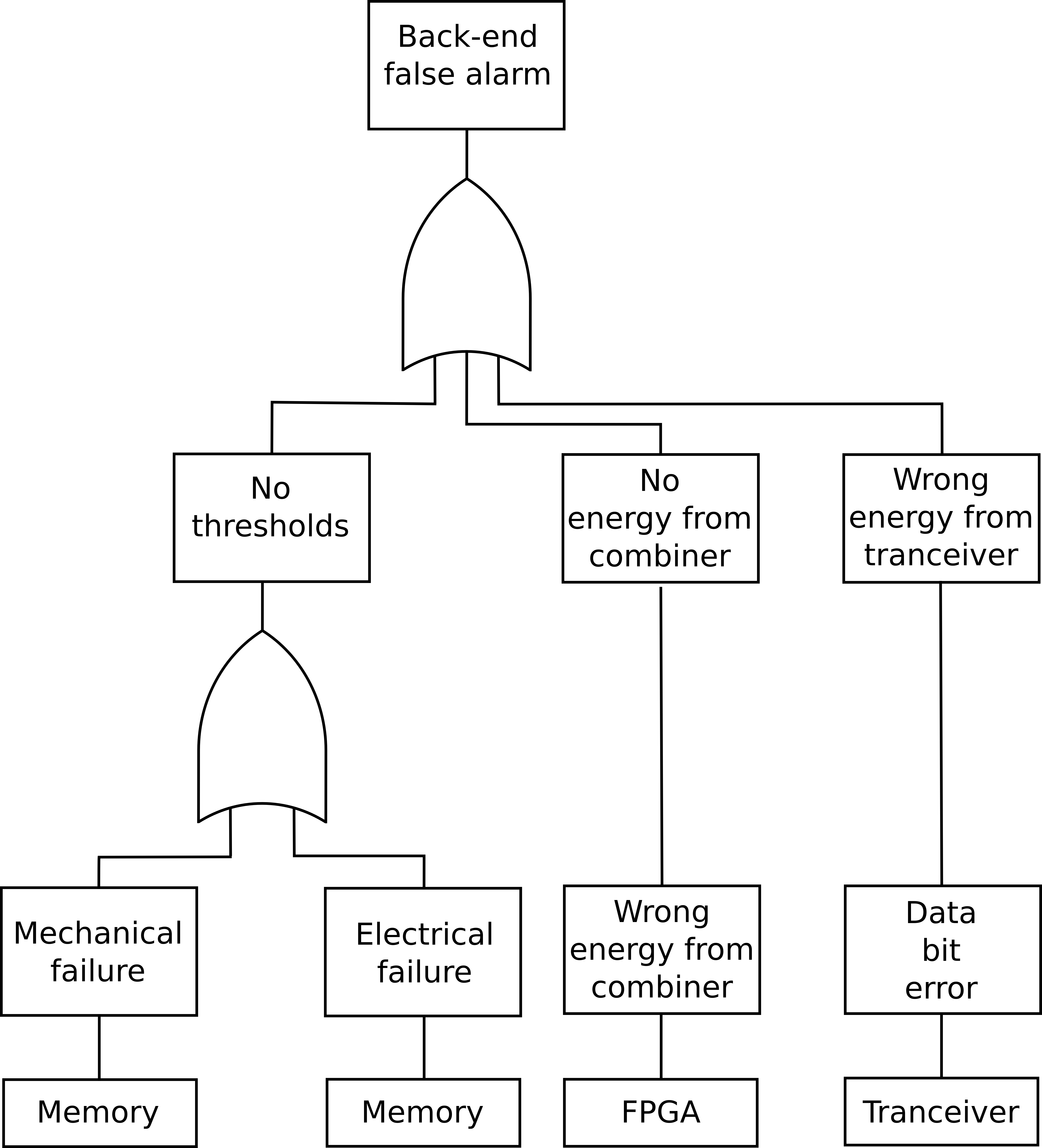}
   \caption{Image section of the false alarm generation fault tree of the LHC BLM system, showing the part describing the back-end acquisition unit.}
   \label{BLM_at_LHCf6}
\end{figure}

The false alarm slice of the fault tree (see Fig.~\ref{BLM_at_LHCf6}) shows the signal chain for different false alarm generators (memory, beam energy from control unit (combiner), and energy transceiver) of the back-end electronics~\cite{isograph_reliability_2013}. The different inputs are linked with a Boolean `OR' gate so that every single input generates, in the same way, a false alarm and, therefore, a downtime of the system and the LHC.

The results of the fault tree analysis have been essential for the design of the hardware and software, especially for the estimates of failure rates of the optical links and the  propagated consequences of it up to the system damage and false rate probabilities. An optimization process has been instigated, to balance the probabilities of damage rates and false alarms. The failure rate calculations also lead to the definition of functional tests and their frequencies. Failure modes are also defined for the limit values, detector names, channel assignments, and much more data needed by the system. Therefore, setting management and metadata verification tests are also treated in the fault tree analysis.

\section{Functionality checks}
\begin{figure}
   \centering
   \includegraphics[width=80mm]{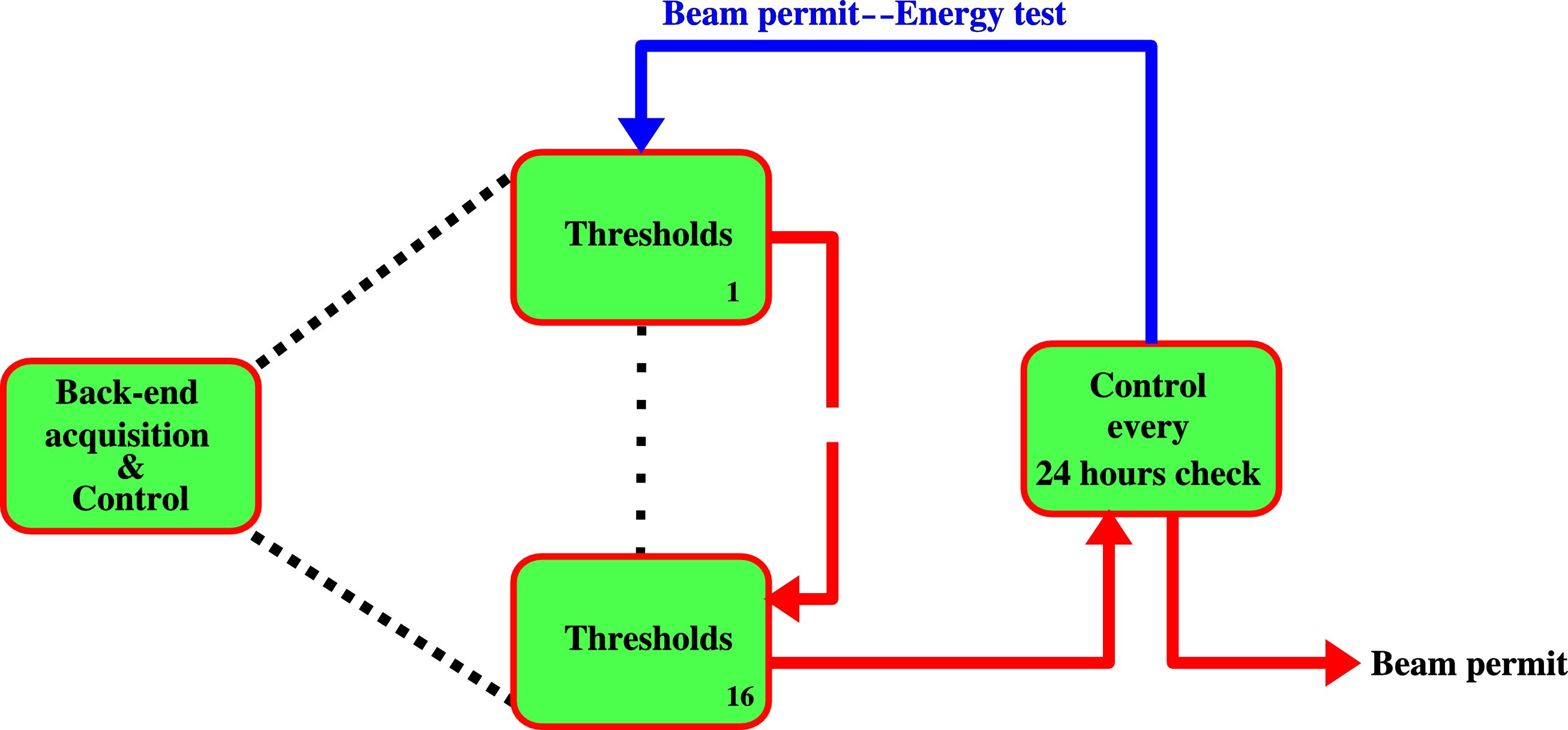}
   \caption{Beam permit line functionality check}
   \label{BLM_at_LHCf7}
\end{figure}

As an example of a check, the signal distribution inside the VME crate for the beam energy and the beam permit line test is discussed~\cite{zamantzas_reliability_2010,dehning_self_2011} (see Fig.~\ref{BLM_at_LHCf7}). The test is initiated by a client, to allow optimal scheduling. The control unit (combiner card) holds a downtime counter requiring every 24 hours the execution of functional tests every 24 hours. If the tests are not completed in time, the downtime counter inhibits the beam permit immediately if no beam is circulating or when the beam present flag becomes false. For the tests, the whole system changes the status to `mode' and, \eg the control units send a request to inhibit the beam permit line to each acquisition card (threshold card) in sequence (see Fig.~\ref{BLM_at_LHCf7}).

\begin{figure}
   \centering
   \includegraphics[width=170mm]{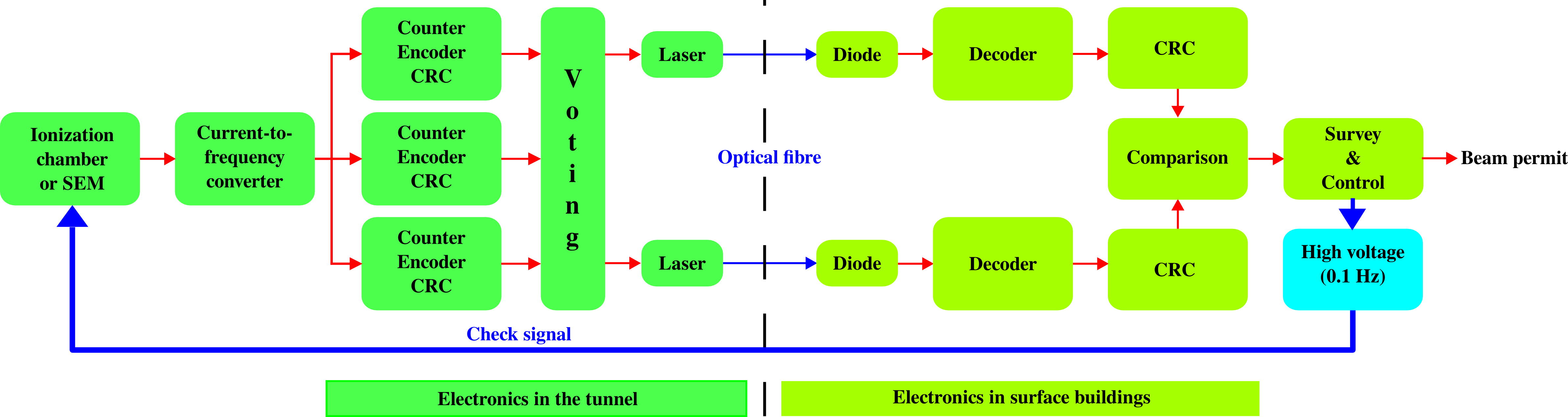}
   \caption{Check of the whole acquisition chain}
   \label{BLM_at_LHCf8}
\end{figure}

The test results are analyzed by the controller; if a false status is detected, a manual intervention is required,  to repair the system before the test can be passed without a false status detected.
The distribution of the beam energy levels between the controller and the acquisition card is tested by changing the energy levels in the test mode; this should result in the acquisition card returning the appropriate threshold settings for comparison with the settings sent.

In a second example, the test of the whole acquisition chain is presented~\cite{emery_first_2008,emery_lhc_2009}. An electrical signal is introduced in the sensor by the capacitive coupling of the sensor electrodes and by a harmonic modulation of the applied high voltage supply (see Fig.~\ref{BLM_at_LHCf8}). This test includes the complete signal chain, except for the ionization process in the ionization chambers and the secondary electron emission in the secondary emission monitor detectors. The conversion of the particle shower to an electrical signal in the detector  is tested every few years with a radiative source placed outside the detector. The long intertest interval for this test is possible because the failure mode of a complete gas exchange with air (ionization chamber) or loss of the vacuum (secondary emission detector) of the detectors will still result in an appropriate signal, without loss of protection functionality.
Also, this test is initiated and the results are analyzed by the back-end unit (survey and control) (see Fig.~\ref{BLM_at_LHCf8}),  allowing  the beam permit line to be inhibited directly in the case of a negative result.

\section{Setting management}
The system setting management controls the settings for the beam permit thresholds and also the settings used for system operation~\cite{nebot_del_busto_handling_2011,holzer_generation_2008}. These operational settings include hard and firmware information, to verify that the configuration stored in the database images the installed system. Table~\ref{table_settings} illustrates the variety of the metadata needed to interpret the measured values or to check the configuration of the system. For example, a match between measured value, channel official names, channel expert names, DCUM (position of monitor), and monitor coefficient needs to be given and tested. To reduce the complexity of the metadata information chain (see Fig.~\ref{BLM_at_LHCf4}, right blocks), a single path is defined for the metadata flow and the measurement values into the back-end unit. The back-end unit distributes the measurement values together with the metadata to ensure consistency and to have only one location where the data integrity needs to be tested. This concept is essential to reduce the number of possible failure modes for metadata corruption.

\begin{table}\scriptsize
        \centering
        \caption{Parameters deployed on each back-end unit (threshold comparator module)}
        \label{table_settings}
        \vspace{3mm}
\begin{tabular}{l r l}
\hline\hline
\textbf{Parameters}                     &\textbf{ Data 32 bit} & \textbf{Description} \\
\hline
Threshold values     &   8192     & 16 channels $\times$ 12 sums $\times$ 32 energies \\
Channel connected       & 1     & Generating (or not) a beam permit \\
Channel mask          & 1           &  `Maskable' or `unmaskable' \\
Serial A     & 1     & Card's serial number (channels 1--8) \\
Serial B       & 1    & Card's serial number (channels 9--16) \\
Serial            & 2     & Threshold comparator \\
Firmware version        &       1        & Threshold comparator's firmware \\
Expert names            & 128           &\\
Official names          & 128           &\\
DCUM                 & 16    & Position of monitor\\
Family names                    & 128   & Threshold family name\\
Monitor coefficients            & 16    & Monitor threshold coefficients\\
Last link-state advertisement update              & 2             & Time stamp: master table\\
Last flash update        & 2         & Time stamp: non-volatile memory \\
Flash checksum       & 1      & CRC value for  or from table integrity\\
\hline\hline
\end{tabular}
\end{table}

\begin{figure}
   \centering
   \includegraphics[width=60mm]{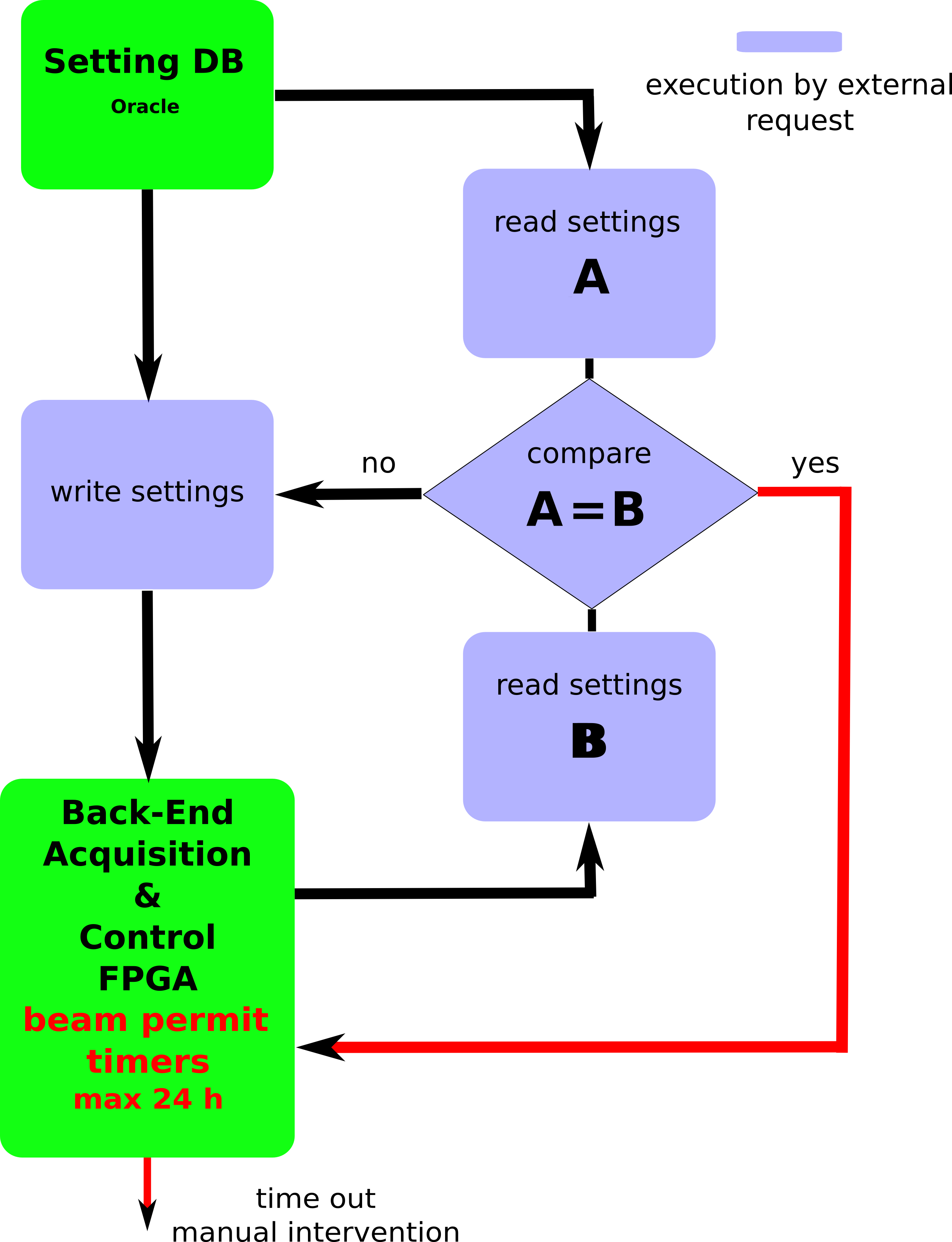}
   \caption{Comparison of descriptive metadatabase reference settings with settings in the back-end acquisition and control unit. The decision logic is indicated in the flow diagram. FPGA,  field-programmable gate array. \textcolor{red}{}}
   \label{BLM_at_LHCf9}
\end{figure}

Having expressed the importance of a failure mode optimized metadata flow, the  data check is achieved by comparing the data stored in a reference setting database (Oracle) with the data stored in the memory of the back-end electronics  field-programmable gate arrays (see Fig.~\ref{BLM_at_LHCf9}). Also, in this test, a downtime counter located in the back-end unit (survey and control) requests a comparison of the data stored at both locations every $24\Uh$. If the test is not initiated, or if the test result is negative, the beam permit is inhibited. Since the comparison is made in a different software environment, the additional functionality required in the back-end unit is marginal, but it is necessary to test the comparison code from time to time.

\subsection{Descriptive metadata}
Metadata need to be generated and the option for required changes needs to be provided. To reduce human error, the graphical user interfaces (GUI) accessing the setting database (see Fig.~\ref{BLM_at_LHCf4}, right block) need to be optimized by allowing for all data manipulation steps to include comparisons with previous data, checks on the magnitudes of changes, and several confirmation steps. The last confirmation steps require the electronic signatures of two independent persons.

The generation of sets of metadata required initially and for larger changes during the operation periods is done for the LHC system by a GUI for the database access. Generation of metadata, such as limits for the beam abort thresholds, are parameterized  and the calculation is made by code loaded into the database (Oracle) (see Fig.~\ref{BLM_at_LHCf4}, rightmost block). The calculation made in the database environment, where database software changes and updates are made in a coherent manner, should ensure long-term maintainability~\cite{nemci_calculation_2012}.

\subsection{Documentation}
In a complex system, designed for operation over decades, sufficient documentation is essential to describe the system for knowledge transfer. For a safety system, the function of the documentation is to avoid  failure modes and failures. The design documentation, from the specification to documentation on operation and system changes, needs to be distributed for review, comment, and final approval by each client. At the LHC, standardized forms, electronic procedures, and signatures are in use to organ\-ize the process, \eg an engineering change request outlines the motivation for a change, the description of the proposed change, and an estimate of the impact of the change on the functionality of the concerned system and other systems.

\section{Snapshots of loss measurements triggered by events}
The loss measurement recording rate has been set up at different speeds, with $40\Uus, 80\Uus, 80\Ums$, and $1.3\Us$ integration times. The two first periods are event-triggered, to cope with the amount of data, while the latter periods are read out at $12\UHz$ and $1\UHz$. The event-triggered measurements are used to analyze losses occurring at particular times during operation or depending on measurements and output analysis data acquisition freezing events. The $12\UHz$ measurements are used for the collimator positioning feedback system and the $1\UHz$ measurements are used for continuous observation of the accelerator status.

\begin{figure}
   \centering
   \includegraphics[width=80mm]{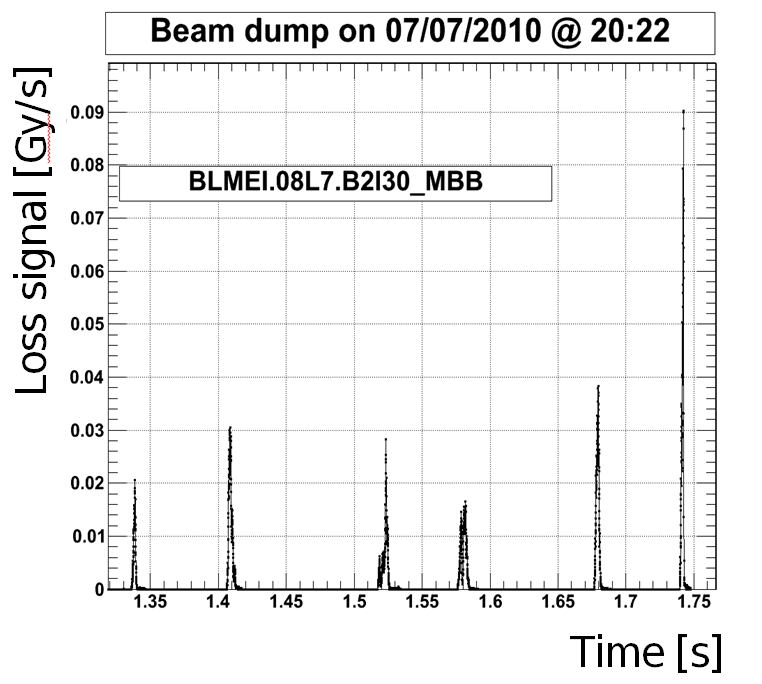}
   \caption{Example of a particle loss triggered event recording. The trigger has been  generated at $1.74\Us$. The measurements recorded before the trigger event reveal loss precursors. The losses are caused by collisions between the beam and dust particles.}
    \label{BLM_at_LHCf10}
\end{figure}

High-resolution data have been used not only for the detailed study of  beam losses caused by dust events (see Fig.~\ref{BLM_at_LHCf10}), but also to check for non-conformities of the acquisition system. When testing the system under extreme conditions, high loss levels with a large leading signal transition give an insight into system performance. The advantage of publishing different measurement signals is that it enables consistency checks to be performed. In the LHC, several clients are used to check the consistency of measurement data.

\section{Acquisition database}
\begin{figure}
   \centering
   \includegraphics[width=80mm]{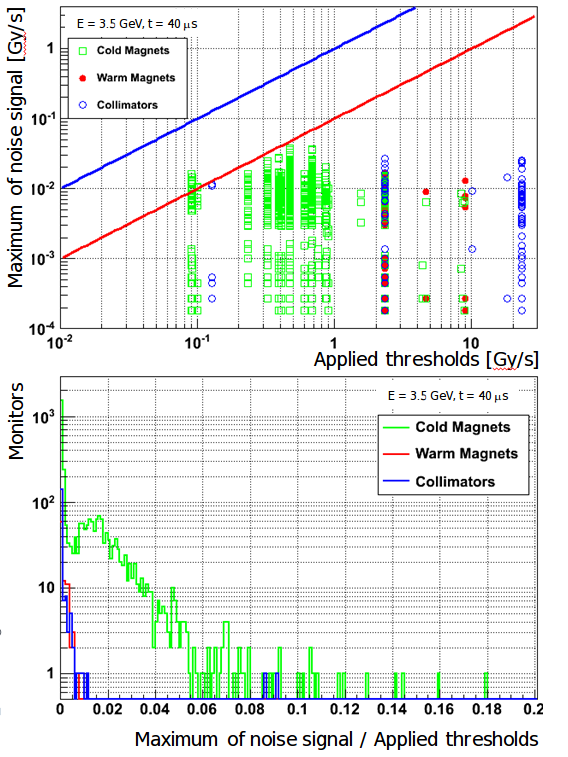}
   \caption{Noise level determination of all beam loss monitor channels. The LHC loss monitor channels are grouped by the observed loss, creating elements of cold and warm magnets and collimators. Top: Beam loss monitor noise signal taken with no beam circulating versus beam abort thresholds. The blue line indicates the threshold value and the red line the maximum noise goal set to avoid any noise false beam aborts. Bottom: Beam loss monitor spectrum normalized to the beam abort threshold.}
      \label{BLM_at_LHCf11}
\end{figure}

The storage and fast retrieval of measurement data and metadata is also essential for system checks. Besides the examples discussed previously, for which extended data storage were required, an extreme case is the check of noise amplitudes of the system (see Fig.~\ref{BLM_at_LHCf11}). For a protection system with limits leading automatically to a beam abort and to accelerator downtime, there is a strong requirement to avoid false aborts caused by rare events (noise). This is extreme, because rare signals need to be retrieved from stored measurement data from acquisition periods lasting weeks. The measurements with the shortest integration periods of $40\Uus$ show the largest signal fluctuation, because signal averaging does not led to a reduction in signal fluctuation. To reduce the amount of data to be stored, an on-line measurement data reduction algorithm has been implemented in the back-end unit. Only maximum values of the short integration times are stored for the $1\UHz$ read-out. This procedure reduces the quantity of data to be stored by over four orders of magnitude. In addition,  a retrieval time optimized database structure has been implemented for this purpose.

\section{Preventive action}
The discussion in Section~\ref{sec:failureprobability} emphasized the reduction in failure rate achieved by surveying the system, to anticipate possible failure modes. In the LHC system, this survey task is realized by the daily retrieval of relevant database information and an automatic comparison with limits for initiating actions. Reports containing different levels of abstraction are produced daily and weekly.
\begin{figure}
   \centering
   \includegraphics[width=80mm]{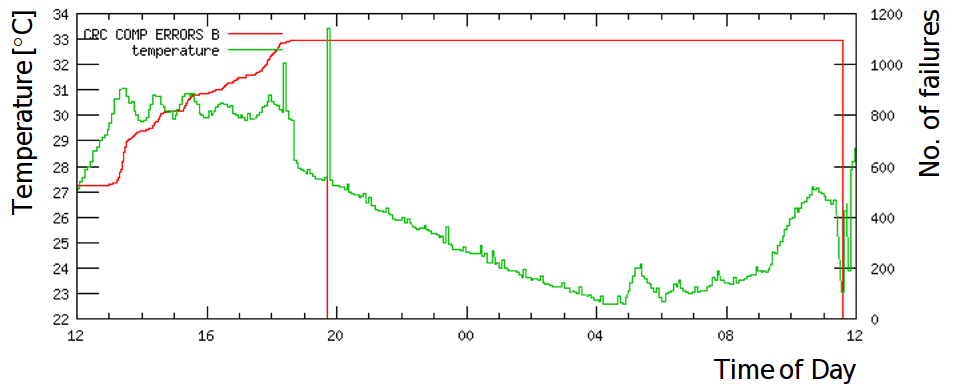}
   \caption{Optical link failures and printed circuit board temperatures versus time of day}
      \label{BLM_at_LHCf12}
\end{figure}
An example of this procedure is given by the survey of optical links. The links are redundant (see Fig.~\ref{BLM_at_LHCf5}) and the calculations of different  CRCs enable the differences between the CRC values to be recorded and correlated with board temperature variations (see Fig.~\ref{BLM_at_LHCf12}). The limits for actions are set empirically, to minimize downtime and maintenance efforts.

\section{Summary}
A systematic design approach for machine protection systems will start with determination of the system failure rate. The failure rate magnitude could be based on well-established standards first developed for the design of military equipment, the aircraft industry, space missions, or nuclear power stations. The effect of increasing complexity by adding protection functionalities and therefore reducing availability is best studied by reliability software packages~\cite{bhattacharyya_reliability_2012}. The basic means of delivering a reduction in failure rate are provided by a system layout with parallel, redundant information, treated in combination with a regular survey of the system status and functional tests. A survey will allow preventive actions, to reduce the failure rate. For a protection system, a fail-safe design is essential so that protection is ensured in the case of a failure.

Functionality checks staged for all levels of the signal treatment are implemented for the LHC BLM system. The checks of the information exchange inside the VME crate and the analogue and digital signal chain have been discussed. Examples have been given to emphasize the importance of the metadata information flow. The combination of measurement and metadata as early as possible in the signal chain is important for the reduction of failure modes and simplified test options.
To attain low failure rates, rigorous metadata tests have to be implemented, to ensure metadata conformity. The generation of metadata and change options using a graphical interface also need to be analyzed in terms of failure modes, taking into account long-term usage and the maintainability of tests and validation pro\-cedures in the future. For the LHC, the most stringent requirement in avoiding human error is the request of two signatures to validate metadata changes. Although listed last, documentation tasks should be started first, including planning for reliability measures, and have to be continued as long as the system exists.

\end{document}